%% file: KG20C-QA.tex
\documentclass[conference]{IEEEtran}
\IEEEoverridecommandlockouts

\usepackage{cite}
\usepackage{amsmath,amssymb,amsfonts}
\usepackage{algorithmic}
\usepackage{graphicx}
\usepackage{textcomp}
\usepackage{xcolor}
\def\BibTeX{{\rm B\kern-.05em{\sc i\kern-.025em b}\kern-.08em
    T\kern-.1667em\lower.7ex\hbox{E}\kern-.125emX}}

\usepackage{lipsum} 
\usepackage[disable]{todonotes} 

\usepackage{graphicx} 
\usepackage{subfigure}
\usepackage{tikz} 
\usepackage{adjustbox}  

\usepackage{booktabs} 
\usepackage{multirow} 
\usepackage{makecell} 
\usepackage{array, tabularx, boldline} 

\usepackage{amsmath} 

\usepackage{amsthm}
\usepackage{amsfonts}
\usepackage{bm} 
\usepackage{mathtools} 
\usepackage{mathdots} 
\usepackage{xfrac} 
\usepackage{faktor} 
\usepackage{cancel} 

\usepackage{algorithm}
\usepackage{url}

\input{./math_commands.tex} 

\begin{document}

\title{KG20C \& KG20C-QA: Scholarly Knowledge Graph Benchmarks for Link Prediction and Question Answering}

\author{\IEEEauthorblockN{Hung-Nghiep Tran}
\IEEEauthorblockA{\textit{\small University of Information Technology, Vietnam} \\
	\textit{\small Vietnam National University, Ho Chi Minh City, Vietnam}\\
	nghiepth@uit.edu.vn}
\and
\IEEEauthorblockN{Atsuhiro Takasu}
\IEEEauthorblockA{\textit{\small National Institute of Informatics, Tokyo, Japan} \\
	\textit{\small The Graduate University for Advanced Studies, SOKENDAI, Japan}\\
	takasu@nii.ac.jp}
}

\maketitle

\begin{abstract}
In this paper, we present KG20C and KG20C-QA, two curated datasets for advancing question answering (QA) research on scholarly data. KG20C is a high-quality scholarly knowledge graph constructed from the Microsoft Academic Graph through targeted selection of venues, quality-based filtering, and schema definition. Although KG20C has been available online in non-peer-reviewed sources such as GitHub repository, this paper provides the first formal, peer-reviewed description of the dataset, including clear documentation of its construction and specifications. KG20C-QA is built upon KG20C to support QA tasks on scholarly data. We define a set of QA templates that convert graph triples into natural language question--answer pairs, producing a benchmark that can be used both with graph-based models such as knowledge graph embeddings and with text-based models such as large language models. We benchmark standard knowledge graph embedding methods on KG20C-QA, analyze performance across relation types, and provide reproducible evaluation protocols. By officially releasing these datasets with thorough documentation, we aim to contribute a reusable, extensible resource for the research community, enabling future work in QA, reasoning, and knowledge-driven applications in the scholarly domain. The full datasets will be released at \url{https://github.com/tranhungnghiep/KG20C/} upon paper publication.
\end{abstract}


\section{Introduction}

Scholarly knowledge graphs (KGs) provide structured representations of the publication ecosystem, capturing entities such as authors, papers, venues, affiliations, and citations. Unlike scientific data, which encodes domain-specific facts (e.g., chemical formulas or biomedical interactions), scholarly data encodes the metadata of the scientific process itself: who wrote which paper, in which venue, under which affiliation, and citing which works. Such information underlies applications including bibliographic search, expert finding, trend analysis, and recommendation systems \cite{sinha_overviewmicrosoftacademic_2015, wang_microsoftacademicgraph_2020}. With the increasing availability of large-scale scholarly graphs such as the Microsoft Academic Graph (MAG) \cite{wang_microsoftacademicgraph_2020}, it becomes important to establish principled benchmark datasets that enable reproducible evaluation of algorithms for reasoning over scholarly metadata.

Benchmark datasets have been central to progress in knowledge graph representation learning. Widely used resources such as WN18RR \cite{dettmers_convolutional2dknowledge_2018, millergeorgea._wordnetlexicaldatabase_1995} and FB15k-237 \cite{toutanova_observedlatentfeatures_2015, bollacker_freebasecollaborativelycreated_2008} have defined standard evaluation settings for link prediction and completion. However, these benchmarks are lexical or encyclopedic in nature and do not reflect the structure or challenges of scholarly metadata and its applications. Meanwhile, large scholarly graphs such as MAG are massive, noisy, and lack standardized train/validation/test splits. Without curated datasets and controlled evaluation settings, it is difficult to ensure fair comparison and reproducibility.

We present \textbf{KG20C}, a curated benchmark scholarly knowledge graph. KG20C is constructed in three stages: (1) data extraction from MAG, focusing on well-defined scholarly entity and relation types; (2) graph construction with careful cleaning and pruning to reduce noise; and (3) dataset splitting into disjoint training, validation, and test sets, following established best practices to avoid data leakage issues observed in earlier benchmarks such as WN18 and FB15k \cite{bordes_translatingembeddingsmodeling_2013}. The result is a high-quality benchmark dataset, directly comparable in scope and usage to WN18RR and FB15k-237 \cite{dettmers_convolutional2dknowledge_2018, toutanova_observedlatentfeatures_2015}, but grounded in the scholarly domain. While preliminary versions of KG20C were informally available, this paper provides its first formal peer-reviewed introduction, ensuring a stable, citable, and reproducible resource.

Building on KG20C, we introduce \textbf{KG20C-QA}, a question answering (QA) benchmark over scholarly data. QA has emerged recently as a central task in the era of large language models, providing a natural means of evaluating reasoning ability over structured and textual resources \cite{weston_aicompletequestionanswering_2015, nakano_webgptbrowserassistedquestionanswering_2021}. In the current stage, KG20C-QA consists of one-hop questions grounded in the KG20C schema. It is released in both natural language and entity--relation forms, enabling evaluation of both text-based methods (e.g., LLMs) and graph-based methods (e.g., knowledge graph embeddings). The current release focuses on one-hop QA, defined using single KG triple templates. While one-hop QA may appear simple, this work provides the foundation of QA benchmark grounded in controlled, high-quality scholarly data and prepare for multi-hop QA and reasoning benchmarks in the future.

Our contributions are as follows:
\begin{itemize}
	\item We construct and officially release \textbf{KG20C}, a curated scholarly knowledge graph derived from MAG, with high-quality entity and relation coverage, careful cleaning, and standardized train/validation/test splits.
	\item We construct and release \textbf{KG20C-QA}, a benchmark dataset for question answering on scholarly data, provided in both natural language and entity--relation forms.
	\item We establish reproducible baseline evaluations for link prediction and QA tasks on these datasets, demonstrating their utility as challenging and standardized benchmarks.
\end{itemize}

By releasing KG20C and KG20C-QA, our goal is to provide standardized resources that enable the community to advance methods for reasoning on scholarly data. We expect these benchmarks to support a broad spectrum of research, from classical knowledge graph embeddings to modern QA with large language models.

\section{Related Work}

\subsection{Knowledge Graph Benchmarks}
Benchmark datasets have been central to advancing representation learning on knowledge graphs. Early resources such as WN18 and FB15k provided lexical and encyclopedic graphs, but were later shown to suffer from test leakage through inverse relations, leading to inflated scores \cite{bordes_translatingembeddingsmodeling_2013}. Their corrected variants, WN18RR and FB15k-237, introduced filtered splits to remove leakage and remain widely used as standard testbeds for link prediction \cite{dettmers_convolutional2dknowledge_2018, toutanova_observedlatentfeatures_2015}. Other encyclopedic graphs such as YAGO \cite{mahdisoltani_yago3knowledgebase_2015} and DBpedia \cite{auer_dbpedianucleusweb_2007} provide rich coverage but lack standardized splits suitable for controlled embedding evaluation.

\subsection{Scholarly Knowledge Graphs}
Several large-scale scholarly graphs have been developed to capture the publication ecosystem, including the Microsoft Academic Graph (MAG) \cite{wang_microsoftacademicgraph_2020, sinha_overviewmicrosoftacademic_2015} and AMiner \cite{tang_arnetminerextractionmining_2008}. These resources span millions of entities and relations across authors, papers, venues, and affiliations, and are widely used for bibliometrics and scientometrics. However, they are massive, noisy, and not released as curated benchmarks with standardized training, validation, and test splits. Without such controlled datasets, reproducibility and fair comparison across methods remain limited.

\subsection{QA Benchmarks on Knowledge Graphs}
Question answering (QA) over knowledge graphs has become an important evaluation paradigm, especially for natural language interfaces. Established QA benchmarks include SimpleQuestions \cite{bordes_largescalesimplequestion_2015}, WebQuestions \cite{berant_semanticparsingfreebase_2013}, MetaQA \cite{zhang_variationalreasoningquestion_2018}, and LC-QuAD \cite{trivedi_lcquadcorpuscomplex_2017}, which provide natural language queries grounded in encyclopedic KGs such as Freebase or DBpedia. These datasets support both single-hop and multi-hop reasoning and have shaped much of the recent progress in KG-based QA. 

Recently, QA datasets have also been developed for scholarly data. SciQA \cite{auer_sciqascientificquestion_2023} introduces questions grounded in the Open Research Knowledge Graph (ORKG), covering scholarly entities and scientific concepts, though it is relatively small in scale and focused on ORKG’s comparison tables. DBLP-QuAD \cite{banerjee_dblpquadquestionanswering_2023} provides one-hop and multi-hop QA benchmarks derived from the DBLP scholarly graph, constructed using query templates and paraphrasing. While highly relevant, DBLP-QuAD is tailored to DBLP and does not align with the rigorous standards of KG completion benchmarks such as WN18RR or FB15k-237.

\subsection{Our Contribution in Context}
Our work provides another curated benchmark resource. KG20C introduces a high-quality scholarly knowledge graph derived from MAG, constructed with careful cleaning and standardized splits that follow best practices to avoid leakage. This yields a benchmark of comparable rigor to WN18RR and FB15k-237, but focused on scholarly metadata. Building on this, KG20C-QA provides a QA benchmark grounded in KG20C, released in both natural language and entity--relation forms. This dual format enables evaluation with both large language models and graph-based methods, while offering a foundation for future multi-hop QA over scholarly data. In contrast to prior QA datasets such as SciQA and DBLP-QuAD, our release emphasizes reproducibility, simplicity of format, and grounding in a curated KG completion benchmark.

\section{The KG20C Dataset}
\label{sec:kg20c}

\subsection{Overview and Motivation}
\begin{figure*}[ht]
	\centering
	\includegraphics[width=0.6\linewidth]{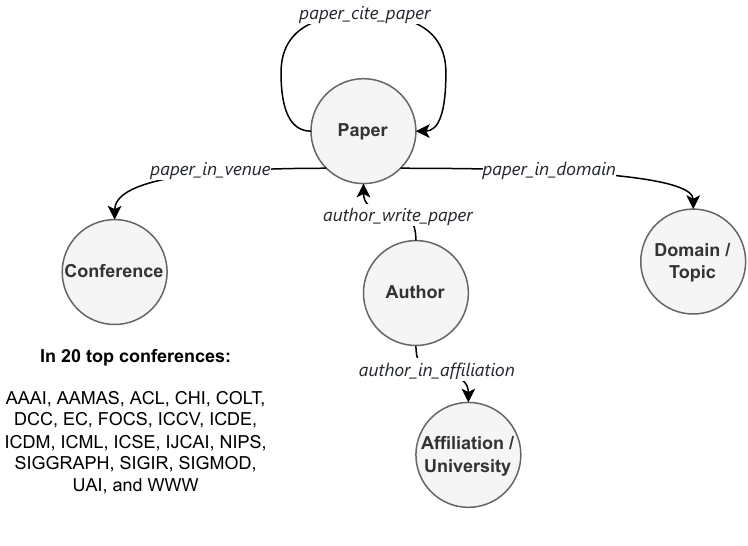}
	\caption{Overview of the KG20C knowledge graph. Circles denote entity types, arrows denote relation types.}
	\label{fig:kg20c_graph}
\end{figure*}

KG20C is a curated scholarly knowledge graph constructed to serve as a benchmark dataset for representation learning and reasoning on scholarly metadata. Unlike encyclopedic benchmarks such as WN18RR or FB15k-237, KG20C is grounded in the publication ecosystem of computer science. The dataset covers twenty top-tier conferences, carefully selected and cleaned to ensure both quality and reproducibility. With standardized splits and a simple file organization, KG20C is designed as a benchmark-ready resource for tasks such as link prediction, knowledge graph embeddings, scholarly recommendation, and question answering over academic content. An overview of the KG20C schema is shown in Figure~\ref{fig:kg20c_graph}.

\subsection{Construction Protocol}

\subsubsection{Scholarly Data Extraction}
KG20C is derived from the Microsoft Academic Graph (MAG) \cite{sinha_overviewmicrosoftacademic_2015}, one of the largest bibliographic datasets available. To ensure quality, we selected papers published between 1990 and 2010 in top-tier computer science venues, identified using the CORE 2020 A* conference ranking. We removed venues with fewer than 300 publications and papers with fewer than 20 recorded citations, filtering out incomplete or low-impact records. The final set covers 20 leading conferences: 
\textit{AAAI, AAMAS, ACL, CHI, COLT, DCC, EC, FOCS, ICCV, ICDE, ICDM, ICML, ICSE, IJCAI, NIPS, SIGGRAPH, SIGIR, SIGMOD, UAI,} and \textit{WWW}. 
This curated bibliographic dataset is referred to as MAG20C. Its descriptive statistics are shown in Table~\ref{tab:data_semquery}.

\setlength{\tabcolsep}{4.5pt}
\begin{table}[ht]
	\centering
	\caption{Data statistics of the curated bibliographic dataset MAG20C.}
	\label{tab:data_semquery}
	\begin{tabular}{@{\extracolsep{-7pt}}lrrrrrrr}
		\toprule
		Dataset & Paper & Author & Affil. & Conf. & Jour. & Domain & Year \\
		\hline
		MAG & 123{,}056{,}983 & 114{,}698{,}044 & 19{,}843 & 1{,}283 & 23{,}404 & 53{,}834 & 1800--2017 \\
		MAG20C & 5{,}047 & 8{,}680 & 692 & 20 & 0 & 1{,}923 & 1990--2010 \\
		\hline
	\end{tabular}
\end{table}

\subsubsection{Knowledge Graph Construction}
From MAG20C, we constructed a labeled multi-digraph with five intrinsic entity types: \textit{Paper}, \textit{Author}, \textit{Affiliation}, \textit{Venue}, and \textit{Domain}. Edges are defined via five intrinsic relation types: \textit{author\_in\_affiliation}, \textit{author\_write\_paper}, \textit{paper\_in\_domain}, \textit{paper\_cite\_paper}, and \textit{paper\_in\_venue}. These relation types are intrinsic, non-redundant, and not directly implied by one another, ensuring that triples cannot be trivially inferred from others.

\subsubsection{Benchmark Data Splitting}
We uniformly split at random the constructed triples into training, validation, and test sets. To ensure fair evaluation, we guaranteed that all entities and relations appearing in the validation and test sets also appear in the training set, so that embeddings can be learned. We made sure that no duplicate triples or inverse-relation leakage were introduced, making KG20C comparable in rigor to WN18RR and FB15k-237. The resulting dataset statistics are summarized in Table~\ref{tab:data_semquery_kg}.

\begin{table}[ht]
	\centering
	\caption{Data statistics of the KG20C knowledge graph.}
	\label{tab:data_semquery_kg}
	\begin{tabular}{@{\extracolsep{5pt}}lrrrrr}
		\toprule
		Dataset & $|\gE|$ & $|\gR|$ & Training & Validation & Test \\
		\hline
		KG20C & 16{,}362 & 5 & 48{,}213 & 3{,}670 & 3{,}724 \\
		\hline
	\end{tabular}
\end{table}

\subsection{Dataset Content and Format}

\subsubsection{File Organization}
The dataset is distributed in plain text, tab-separated-values (TSV) format, consistent with established KG benchmarks such as WN18RR and FB15k-237, making it ready to be used by many KG embedding libraries. Core files include:
\begin{itemize}
	\item \textit{all\_entity\_info.txt}: entity ID, name, and type.
	\item \textit{all\_relation\_info.txt}: list of relation IDs.
	\item \textit{train.txt}, \textit{valid.txt}, \textit{test.txt}: triples of the form \textit{entity\_id \;\; relation\_id \;\; entity\_id}.
\end{itemize}
For example, the entry \textit{28674CFA \;\; author\_in\_affiliation \;\; 075CFC38} denotes that the author with ID \textit{28674CFA} is affiliated with institution \textit{075CFC38}.

\subsubsection{Dataset Statistics}
KG20C contains 16,362 entities connected by 48,000+ training triples and nearly 7,400 validation and test triples. Its scale and relational structure make it comparable to high-quality standard link prediction benchmarks, especially WN18RR, as shown in Table~\ref{tab:data}. 

\begin{table}[ht]
	\centering
	\caption{Statistics of standard benchmarks for comparison.}
	\label{tab:data}
	\begin{adjustbox}{max width=\columnwidth}
		\begin{tabular}{@{\extracolsep{0pt}}lrrrrr}
			\toprule
			Dataset & $|\gE|$ & $|\gR|$ & Train & Valid & Test \\
			\hline
			WN18RR & 40{,}943 & 11 & 86{,}835 & 3{,}034 & 3{,}134 \\
			FB15K-237 & 14{,}541 & 237 & 272{,}115 & 17{,}535 & 20{,}466 \\
			\hline
		\end{tabular}
	\end{adjustbox}
\end{table}

\subsection{Positioning and Contribution}
KG20C extends the family of high-quality link prediction benchmarks to the scholarly domain. Its standardized splits, curated scope, and simple format ensure reproducibility and comparability across methods. While encyclopedic datasets such as WN18RR and FB15k-237 remain dominant in benchmarking, KG20C provides an equally rigorous resource grounded in scholarly metadata. By bridging bibliographic information with KG benchmarks, KG20C establishes a foundation for evaluating embeddings, link prediction, recommendation, and question answering in scientific contexts.

\section{The KG20C-QA Dataset}
\label{sec:kg20cqa}

\subsection{Motivation}
While knowledge graph evaluation is traditionally carried out via link prediction tasks, many real-world information needs are naturally expressed in the form of questions. For instance, a researcher may ask \textit{Who wrote this paper?} or \textit{What conferences published this work?} Such queries highlight the importance of a question answering (QA) perspective on scholarly knowledge graphs. 

To bridge the gap between graph-based benchmarks and natural language understanding, we extend KG20C into a question answering dataset, called \textbf{KG20C-QA}. This resource enables evaluation of both graph embedding methods and natural language models. In doing so, KG20C-QA makes scholarly data analysis more accessible across the knowledge graph and NLP communities, complementing existing QA benchmarks such as SimpleQuestions or WebQuestions, but grounded in a curated scholarly domain.

\begin{table*}[ht]
	\small
	\centering
	\caption{Question templates for KG20C-QA, by relation type and query direction.}
	\label{tab:qa_templates}
	\begin{tabular}{p{3cm}p{5.5cm}p{6.5cm}}
		\toprule
		\textbf{Relation} & \textbf{Forward query (head $\to$ tail)} & \textbf{Reverse query (tail $\to$ head)} \\
		\hline
		author\_in\_affiliation & Where does this author work? & Who works at this organization? \\
		author\_write\_paper & What papers did this author write? & Who wrote this paper? \\
		paper\_in\_domain & What domains does this paper belong to? & What papers belong to this domain? \\
		paper\_cite\_paper & Which papers does this paper cite? & Which papers cite this paper? \\
		paper\_in\_venue & Where was this paper published? & What papers were published in this conference? \\
		\bottomrule
	\end{tabular}
\end{table*}

\subsection{Task Definition}
Each triple $(h, r, t)$ in KG20C is reframed into one or more natural language question--answer pairs. For each relation type, we define two query directions:
\begin{itemize}
	\item \textbf{Forward queries}: Given the head entity and relation, predict the tail entity. 
	\item \textbf{Reverse queries}: Given the tail entity and relation, predict the head entity.
\end{itemize}
Thus, every triple produces a pair of question, expressed in:
\begin{enumerate}
	\item \textit{Entity--relation form}, i.e., incomplete triples such as $(h, r, ?)$.
	\item \textit{Natural language form}, i.e., template-based questions such as \textit{What papers did this author write?}.
\end{enumerate}

\subsection{Construction Protocol}
We designed question templates for each of the five intrinsic relation types in KG20C. Each template produces natural, human-readable questions that avoid ambiguity. For example, the relation \textit{author\_write\_paper} is expressed as either \textit{What papers did this author write?} (forward) or \textit{Who wrote this paper?} (reverse). Note that the names of the entities (i.e., author name, paper title, etc.) are used to form complete natural language questions.

The resulting dataset is released in two synchronized forms:
\begin{itemize}
	\item \textbf{Graph form}: triples in TSV format, compatible with embedding-based QA and link prediction.
	\item \textbf{Text form}: natural language questions with corresponding entity answers, supporting text-based QA models.
\end{itemize}

\subsection{Dataset Content and Statistics}
By systematically applying these templates, every triple in KG20C corresponds to two QA pairs. In total, KG20C-QA provides a large-scale yet controlled QA benchmark, with standardized train/valid/test splits aligned with KG20C. The data statistics are shown in Table~\ref{tab:data_kg20c-qa}. Example queries for each relation type are shown in Table~\ref{tab:qa_templates}.

\begin{table}[ht]
	\centering
	\caption{Data statistics of the KG20C-QA dataset.}
	\label{tab:data_kg20c-qa}
	\begin{tabular}{@{\extracolsep{5pt}}lrrrrr}
		\toprule
		Dataset & Training & Validation & Test \\
		\hline
		KG20C-QA & 96{,}426 & 7{,}340 & 7{,}448 \\
		\hline
	\end{tabular}
\end{table}
\todo{Compute data statistics for each question type: count how many question of each type.}

\subsection{Positioning and Contribution}
KG20C-QA is designed for dual compatibility:
\begin{itemize}
	\item \textbf{Graph-based QA models}: treat QA as triple completion, using embedding methods or link prediction approaches.
	\item \textbf{Text-based QA models}: input natural language questions and output the correct entity, aligning with the NLP community’s QA tasks.
\end{itemize}
This dual nature makes KG20C-QA a versatile testbed for evaluating the intersection of knowledge graphs and language models. While the present release focuses on one-hop questions, it establishes a foundation for future multi-hop QA benchmarks in the scholarly domain.

KG20C-QA complements KG20C by adding a question answering layer. It provides a reproducible, curated, and extensible benchmark that connects knowledge graph embeddings with natural language QA. Compared to encyclopedic QA datasets, KG20C-QA emphasizes high-quality scholarly data, standardized splits, and alignment with a curated KG benchmark. We expect it to foster research across both KG and NLP communities on real-world scholarly reasoning tasks.

\section{Baseline Benchmarking Experiments} \label{sect:baseline}

In this section, we present baseline experiments on KG20C and KG20C-QA datasets for multiple strong models. We describe the experimental setup, training procedure, and evaluation metrics, followed by detailed results on both link prediction and QA-style queries.

\subsection{Experimental Setup}

\paragraph*{Tasks} 
We evaluate models on two tasks: (1) the standard \textit{link prediction} task on KG20C, where given a head (or tail) entity and relation, the goal is to rank candidate tail (or head) entities; and (2) a \textit{QA} task on KG20C-QA, where we answer the questions (e.g., ``What papers may this author write?'') in the entity-relation form using graph-based methods.

\paragraph*{Models} 
We benchmark the following models:
\begin{itemize}
	\item \textbf{Random baseline}: selects the answer uniformly at random among all entities.
	\item \textbf{Word2vec skipgram}: a single-relational embedding baseline trained on co-occurrence windows, based on word2vec \cite{mikolov_distributedrepresentationswords_2013, mikolov_efficientestimationword_2013}.
	\item \textbf{CP$_h$}: a canonical polyadic tensor factorization baseline for multi-relational embedding, as a strong baseline for KG embedding \cite{lacroix_canonicaltensordecomposition_2018}.
	\item \textbf{MEI}: a multi-relational embedding model with matrix-enhanced interactions, as a strong recent baseline for KG embedding \cite{tran_multipartitionembeddinginteraction_2020, tran_meimmultipartitionembedding_2022, tran_multirelationalembeddingknowledge_2020}.
\end{itemize}

\paragraph*{Training protocol}
All models are trained with mini-batch stochastic gradient descent using the Adam optimizer \cite{kingma_adammethodstochastic_2015}. We use full softmax cross-entropy loss with both \textit{1-vs-all} and \textit{k-vs-all} negative sampling \cite{dettmers_convolutional2dknowledge_2018, lacroix_canonicaltensordecomposition_2018}. 
Hyperparameters (weight decay, dropout) are tuned via random search \cite{bergstra_randomsearchhyperparameter_2012}. Embedding sizes are fixed across models as $100$ effective dimensions to keep parameter counts comparable: MEI uses a $10 \times 10$ partitioned embedding, while CP$_h$ and word2vec use $50 \times 2$. 
Training is early-stopped based on validation MRR. All results are the median across three random seeds.

\paragraph*{Evaluation metrics}
We compute the mean reciprocal rank (MRR) and Hits@$k$ for $k \in \{1, 3, 10\}$, following the standard filtered protocol \cite{bordes_translatingembeddingsmodeling_2013}. In addition, we report \textit{type-filtered} metrics, where candidate entities are restricted to the correct semantic type, reflecting realistic use cases.

\subsection{Link Prediction Results}

Table~\ref{tab:result_kg20c} presents results on the standard link prediction benchmark. As expected, random guessing performs near zero. Word2vec, while stronger, remains far below multi-relational embeddings. CP$_h$ significantly outperforms word2vec, while MEI achieves the best overall performance, highlighting its superior expressiveness and parameter efficiency.

\begin{table}[htb]
	\small
	\centering
	\caption{Link prediction results on KG20C.}
	\label{tab:result_kg20c}
	\begin{tabular}{lcccc}
		\toprule
		\textbf{Models} & \textbf{MRR} & \textbf{Hit@1} & \textbf{Hit@3} & \textbf{Hit@10}\\ 
		\hline
		Random & 0.001 & $ < $ 5e-4 & $ < $ 5e-4 & $ < $ 5e-4\\
		Word2vec & 0.068 & 0.011 & 0.070 & 0.177\\
		CP$_h$ & 0.215 & 0.148 & 0.234 & 0.348\\
		MEI & \textbf{0.230} & \textbf{0.157} & \textbf{0.258} & \textbf{0.368}\\
		\bottomrule
	\end{tabular}
\end{table}

%

\subsection{Question Answering Results}

We now solve the \textit{QA} task on KG20C-QA, where we answer the questions (e.g., ``What papers may this author write?'') in the entity--relation form (author, author\_write\_paper, ?) using graph-based methods. Note that experiments with answering the questions in natural language form are left for future work.

Table~\ref{tab:result_kg20c_detail_intype} reports per-query performance for the MEI model under type-filtered evaluation. Results reveal substantial variation across relation types: questions about author--organization or paper--conference yield high accuracy, while domain-related questions remain challenging. These patterns highlight where current embeddings succeed and where further modeling innovations are required.

\begin{table*}[htb]
	\small
	\centering
	\caption{Detailed relational query results with MEI on KG20C-QA (type-filtered).}
	\label{tab:result_kg20c_detail_intype}
	\begin{tabular}{lcccc}
		\toprule
		\textbf{Queries} & \textbf{MRR} & \textbf{Hit@1} & \textbf{Hit@3} & \textbf{Hit@10}\\ 
		\hline
		Who may work at this organization? & 0.299 & 0.221 & 0.342 & 0.440\\
		Where may this author work at? & 0.626 & 0.562 & 0.669 & 0.731\\
		Who may write this paper? & 0.247 & 0.164 & 0.283 & 0.405\\
		What papers may this author write? & 0.273 & 0.182 & 0.324 & 0.430\\
		Which papers may cite this paper? & 0.116 & 0.033 & 0.120 & 0.290\\
		Which papers may this paper cite? & 0.193 & 0.097 & 0.225 & 0.404\\
		Which papers may belong to this domain? & 0.052 & 0.025 & 0.049 & 0.100\\
		Which may be the domains of this paper? & 0.189 & 0.114 & 0.206 & 0.333\\
		Which papers may publish in this conference? & 0.148 & 0.084 & 0.168 & 0.257\\
		Which conferences may this paper publish in? & 0.693 & 0.542 & 0.810 & 0.976\\
		\bottomrule
	\end{tabular}
\end{table*}

\subsection{Summary of Findings}

Across both link prediction and question answering results, we observe:
\begin{itemize}
	\item Multi-relational embeddings (CP$_h$, MEI) strongly outperform single-relational word2vec. Among them, MEI consistently yields the best results, demonstrating its expressiveness and efficiency. 
	\item The benchmark datasets KG20C and KG20C-QA are difficult and not saturated, as indicated by the generally low results of the baselines. These results present them as promising benchmark for the link prediction, QA, reasoning, and other tasks \cite{tran_analyzingknowledgegraph_2019, tran_exploringscholarlydata_2019, tran_encodingsearchingseparationperspective_2024}.
\end{itemize}

\section{Conclusion} \label{sect:conclusion}

We introduced \textbf{KG20C}, a curated scholarly knowledge graph covering 20 leading computer science conferences, and \textbf{KG20C-QA}, a QA benchmark grounded in KG20C. Both datasets are constructed with rigorous filtering, standardized data splits that avoid leakage, and formats aligned with established benchmarks such as WN18RR and FB15k-237. KG20C supports link prediction and embedding evaluation, while KG20C-QA provides both natural language and entity--relation forms, enabling comparative study of graph-based and text-based question answering.

Baseline experiments with random, word2vec, CP$_h$, and MEI methods confirm that these benchmarks remain challenging. MEI performs best, yet overall results are far from saturated. Relation-level QA results further highlight strengths (author--organization, paper--conference) and weaknesses (domain, citation) of current graph-based methods, pointing to open challenges for future modeling.

KG20C and KG20C-QA prioritize reproducibility, standard practices, and benchmark quality compared to broader but noisier resources (e.g., DBLP-derived datasets). While the current release only emphasizes one-hop queries, and the provided baselines only include graph-based methods, we establish a foundation for multi-hop reasoning extension and large language model evaluation in future work.


\bibliographystyle{IEEEtran}
\input{KG20C-QA.bbl}

\end{document}

%% file: math_commands.tex
\def\eqref#1{equation~\ref{#1}}

\def\1{\bm{1}}

\DeclareMathAlphabet{\mathsfit}{\encodingdefault}{\sfdefault}{m}{sl}
\SetMathAlphabet{\mathsfit}{bold}{\encodingdefault}{\sfdefault}{bx}{n}

\def\gE{{\mathcal{E}}}

\def\gR{{\mathcal{R}}}

%% file: KG20C-QA.bbl

%% file: KG20C-QA.bbl
\begin{thebibliography}{10}
\providecommand{\url}[1]{#1}
\csname url@samestyle\endcsname
\providecommand{\newblock}{\relax}
\providecommand{\bibinfo}[2]{#2}
\providecommand{\BIBentrySTDinterwordspacing}{\spaceskip=0pt\relax}
\providecommand{\BIBentryALTinterwordstretchfactor}{4}
\providecommand{\BIBentryALTinterwordspacing}{\spaceskip=\fontdimen2\font plus
\BIBentryALTinterwordstretchfactor\fontdimen3\font minus
  \fontdimen4\font\relax}
\providecommand{\BIBforeignlanguage}[2]{{%
\expandafter\ifx\csname l@#1\endcsname\relax
\typeout{** WARNING: IEEEtran.bst: No hyphenation pattern has been}%
\typeout{** loaded for the language `#1'. Using the pattern for}%
\typeout{** the default language instead.}%
\else
\language=\csname l@#1\endcsname
\fi
#2}}
\providecommand{\BIBdecl}{\relax}
\BIBdecl

\bibitem{sinha_overviewmicrosoftacademic_2015}
A.~Sinha, Z.~Shen, Y.~Song, H.~Ma, D.~Eide, B.-J.~P. Hsu, and K.~Wang, ``An
  {{Overview}} of {{Microsoft Academic Service}} ({{MAS}}) and
  {{Applications}},'' 2015, pp. 243--246.

\bibitem{wang_microsoftacademicgraph_2020}
K.~Wang, Z.~Shen, C.~Huang, C.-H. Wu, Y.~Dong, and A.~Kanakia, ``Microsoft
  {{Academic Graph}}: {{When}} experts are not enough,'' \emph{Quantitative
  Science Studies}, vol.~1, no.~1, pp. 396--413, Feb. 2020.

\bibitem{dettmers_convolutional2dknowledge_2018}
T.~Dettmers, P.~Minervini, P.~Stenetorp, and S.~Riedel, ``Convolutional {{2D
  Knowledge Graph Embeddings}},'' in \emph{Proceedings of the 32nd {{AAAI
  Conference}} on {{Artificial Intelligence}}}, 2018, pp. 1811--1818.

\bibitem{millergeorgea._wordnetlexicaldatabase_1995}
{Miller, George A.}, ``{{WordNet}}: A {{Lexical Database}} for {{English}},''
  \emph{Communications of the ACM}, vol.~38, no.~11, pp. 39--41, 1995.

\bibitem{toutanova_observedlatentfeatures_2015}
K.~Toutanova and D.~Chen, ``Observed versus latent features for knowledge base
  and text inference,'' in \emph{Proceedings of the 3rd {{Workshop}} on
  {{Continuous Vector Space Models}} and Their {{Compositionality}}}, 2015, pp.
  57--66.

\bibitem{bollacker_freebasecollaborativelycreated_2008}
K.~Bollacker, C.~Evans, P.~Paritosh, T.~Sturge, and J.~Taylor, ``Freebase: A
  {{Collaboratively Created Graph Database}} for {{Structuring Human
  Knowledge}},'' in \emph{Proceedings of the 2008 {{ACM SIGMOD International
  Conference}} on {{Management}} of {{Data}}}, 2008, pp. 1247--1250.

\bibitem{bordes_translatingembeddingsmodeling_2013}
A.~Bordes, N.~Usunier, A.~{Garcia-Duran}, J.~Weston, and O.~Yakhnenko,
  ``Translating {{Embeddings}} for {{Modeling Multi-Relational Data}},'' in
  \emph{Advances in {{Neural Information Processing Systems}}}, 2013, pp.
  2787--2795.

\bibitem{weston_aicompletequestionanswering_2015}
J.~Weston, A.~Bordes, S.~Chopra, A.~M. Rush, B.~{van Merri{\"e}nboer},
  A.~Joulin, and T.~Mikolov, ``Towards {{AI-Complete Question Answering}}: {{A
  Set}} of {{Prerequisite Toy Tasks}},'' \emph{arXiv:1502.05698 [cs, stat]},
  Feb. 2015.

\bibitem{nakano_webgptbrowserassistedquestionanswering_2021}
R.~Nakano, J.~Hilton, S.~Balaji, J.~Wu, L.~Ouyang, C.~Kim, C.~Hesse, S.~Jain,
  V.~Kosaraju, W.~Saunders, X.~Jiang, K.~Cobbe, T.~Eloundou, G.~Krueger,
  K.~Button, and M.~Knight, ``{{WebGPT}}: {{Browser-assisted}}
  question-answering with human feedback,'' p.~30, Dec. 2021.

\bibitem{mahdisoltani_yago3knowledgebase_2015}
F.~Mahdisoltani, J.~Biega, and F.~M. Suchanek, ``{{YAGO3}}: {{A Knowledge
  Base}} from {{Multilingual Wikipedias}},'' in \emph{Proceedings of the
  {{Conference}} on {{Innovative Data Systems Research}}}, 2015.

\bibitem{auer_dbpedianucleusweb_2007}
S.~Auer, C.~Bizer, G.~Kobilarov, J.~Lehmann, R.~Cyganiak, and Z.~Ives,
  ``{{DBpedia}}: {{A Nucleus}} for a {{Web}} of {{Open Data}},'' in \emph{The
  {{Semantic Web}}}, D.~Hutchison, T.~Kanade, J.~Kittler, J.~M. Kleinberg,
  F.~Mattern, J.~C. Mitchell, M.~Naor, O.~Nierstrasz, C.~Pandu~Rangan,
  B.~Steffen, M.~Sudan, D.~Terzopoulos, D.~Tygar, M.~Y. Vardi, G.~Weikum,
  K.~Aberer, K.-S. Choi, N.~Noy, D.~Allemang, K.-I. Lee, L.~Nixon, J.~Golbeck,
  P.~Mika, D.~Maynard, R.~Mizoguchi, G.~Schreiber, and P.~{Cudr{\'e}-Mauroux},
  Eds.\hskip 1em plus 0.5em minus 0.4em\relax Berlin, Heidelberg: Springer
  Berlin Heidelberg, 2007, vol. 4825, pp. 722--735.

\bibitem{tang_arnetminerextractionmining_2008}
J.~Tang, J.~Zhang, L.~Yao, J.~Li, L.~Zhang, and Z.~Su, ``{{ArnetMiner}}:
  Extraction and mining of academic social networks,'' in \emph{Proceedings of
  the 14th {{ACM SIGKDD}} International Conference on {{Knowledge}} Discovery
  and Data Mining - {{KDD}} 08}.\hskip 1em plus 0.5em minus 0.4em\relax Las
  Vegas, Nevada, USA: ACM Press, 2008, p. 990.

\bibitem{bordes_largescalesimplequestion_2015}
A.~Bordes, N.~Usunier, S.~Chopra, and J.~Weston, ``Large-scale {{Simple
  Question Answering}} with {{Memory Networks}},'' Jun. 2015.

\bibitem{berant_semanticparsingfreebase_2013}
J.~Berant, A.~Chou, R.~Frostig, and P.~Liang, ``Semantic {{Parsing}} on
  {{Freebase}} from {{Question-Answer Pairs}},'' in \emph{Proceedings of the
  2013 {{Conference}} on {{Empirical Methods}} in {{Natural Language
  Processing}}}, D.~Yarowsky, T.~Baldwin, A.~Korhonen, K.~Livescu, and
  S.~Bethard, Eds.\hskip 1em plus 0.5em minus 0.4em\relax Seattle, Washington,
  USA: Association for Computational Linguistics, Oct. 2013, pp. 1533--1544.

\bibitem{zhang_variationalreasoningquestion_2018}
Y.~Zhang, H.~Dai, Z.~Kozareva, A.~Smola, and L.~Song, ``Variational
  {{Reasoning}} for {{Question Answering With Knowledge Graph}},''
  \emph{Proceedings of the AAAI Conference on Artificial Intelligence},
  vol.~32, no.~1, Apr. 2018.

\bibitem{trivedi_lcquadcorpuscomplex_2017}
P.~Trivedi, G.~Maheshwari, M.~Dubey, and J.~Lehmann, ``{{LC-QuAD}}: {{A
  Corpus}} for {{Complex Question Answering}} over {{Knowledge Graphs}},'' in
  \emph{The {{Semantic Web}} -- {{ISWC}} 2017}, C.~{d'Amato}, M.~Fernandez,
  V.~Tamma, F.~Lecue, P.~{Cudr{\'e}-Mauroux}, J.~Sequeda, C.~Lange, and
  J.~Heflin, Eds.\hskip 1em plus 0.5em minus 0.4em\relax Cham: Springer
  International Publishing, 2017, vol. 10588, pp. 210--218.

\bibitem{auer_sciqascientificquestion_2023}
S.~Auer, D.~A.~C. Barone, C.~Bartz, E.~G. Cortes, M.~Y. Jaradeh, O.~Karras,
  M.~Koubarakis, D.~Mouromtsev, D.~Pliukhin, D.~Radyush, I.~Shilin, M.~Stocker,
  and E.~Tsalapati, ``The {{SciQA Scientific Question Answering Benchmark}} for
  {{Scholarly Knowledge}},'' \emph{Scientific Reports}, vol.~13, no.~1, p.
  7240, May 2023.

\bibitem{banerjee_dblpquadquestionanswering_2023}
D.~Banerjee, S.~Awale, R.~Usbeck, and C.~Biemann, ``{{DBLP-QuAD}}: {{A Question
  Answering Dataset}} over the {{DBLP Scholarly Knowledge Graph}},'' Mar. 2023.

\bibitem{mikolov_distributedrepresentationswords_2013}
T.~Mikolov, I.~Sutskever, K.~Chen, G.~S. Corrado, and J.~Dean, ``Distributed
  {{Representations}} of {{Words}} and {{Phrases}} and {{Their
  Compositionality}},'' in \emph{Advances in {{Neural Information Processing
  Systems}}}, 2013, pp. 3111--3119.

\bibitem{mikolov_efficientestimationword_2013}
T.~Mikolov, K.~Chen, G.~Corrado, and J.~Dean, ``Efficient {{Estimation}} of
  {{Word Representations}} in {{Vector Space}},'' in \emph{Workshop
  {{Proceedings}} of the 2013 {{International Conference}} on {{Learning
  Representations}}}, 2013, p.~12.

\bibitem{lacroix_canonicaltensordecomposition_2018}
T.~Lacroix, N.~Usunier, and G.~Obozinski, ``Canonical {{Tensor Decomposition}}
  for {{Knowledge Base Completion}},'' in \emph{International {{Conference}} on
  {{Machine Learning}}}, 2018, pp. 2863--2872.

\bibitem{tran_multipartitionembeddinginteraction_2020}
H.-N. Tran and A.~Takasu, ``Multi-{{Partition Embedding Interaction}} with
  {{Block Term Format}} for {{Knowledge Graph Completion}},'' in
  \emph{Proceedings of the {{European Conference}} on {{Artificial
  Intelligence}}}, 2020, pp. 833--840.

\bibitem{tran_meimmultipartitionembedding_2022}
------, ``{{MEIM}}: {{Multi-partition Embedding Interaction Beyond Block Term
  Format}} for {{Efficient}} and {{Expressive Link Prediction}},'' in
  \emph{Proceedings of the {{Thirty-First International Joint Conference}} on
  {{Artificial Intelligence}}}, 2022, pp. 2262--2269.

\bibitem{tran_multirelationalembeddingknowledge_2020}
H.-N. Tran, ``Multi-{{Relational Embedding}} for {{Knowledge Graph
  Representation}} and {{Analysis}},'' Ph.D. dissertation, The Graduate
  University for Advanced Studies, SOKENDAI, Japan, 2020.

\bibitem{kingma_adammethodstochastic_2015}
D.~P. Kingma and J.~Ba, ``Adam: {{A Method}} for {{Stochastic Optimization}},''
  in \emph{International {{Conference}} on {{Learning Representations}}}, 2015,
  p.~15.

\bibitem{bergstra_randomsearchhyperparameter_2012}
J.~Bergstra and Y.~Bengio, ``Random {{Search}} for {{Hyper-Parameter
  Optimization}},'' \emph{Journal of Machine Learning Research}, vol.~13, pp.
  281--305, 2012.

\bibitem{tran_analyzingknowledgegraph_2019}
H.-N. Tran and A.~Takasu, ``Analyzing {{Knowledge Graph Embedding Methods}}
  from a {{Multi-Embedding Interaction Perspective}},'' in \emph{Proceedings of
  the {{Data Science}} for {{Industry}} 4.0 {{Workshop}} at {{EDBT}}/{{ICDT}}},
  2019, p.~7.

\bibitem{tran_exploringscholarlydata_2019}
------, ``Exploring {{Scholarly Data}} by {{Semantic Query}} on {{Knowledge
  Graph Embedding Space}},'' in \emph{Proceedings of the 23rd {{International
  Conference}} on {{Theory}} and {{Practice}} of {{Digital Libraries}}}, 2019,
  pp. 154--162.

\bibitem{tran_encodingsearchingseparationperspective_2024}
H.-N. Tran, A.~Aizawa, and A.~Takasu, ``An {{Encoding--Searching Separation
  Perspective}} on {{Bi-Encoder Neural Search}},'' arXiv:2408.01094, Aug. 2024.

\end{thebibliography}
